\documentclass[journal,twoside,web]{ieeecolor}
\usepackage{tuffc}
\usepackage{cite}
\let\labelindent\relax
\usepackage{enumitem}
\usepackage{amsmath,amssymb,amsfonts}
\usepackage{algorithmic}
\usepackage{graphicx}
\usepackage{hyperref}
\usepackage{url}
\usepackage{textcomp}
\usepackage{wrapfig,colortbl}
\usepackage{graphicx} 
\usepackage{graphicx}
\usepackage[rgb]{xcolor}

\usepackage{textcomp}
\usepackage{soul}
\usepackage{svg}
\usepackage{url}
\usepackage{multirow}
\usepackage{makecell}
\usepackage{dblfloatfix}    

\definecolor{abstractbg}{rgb}{1,0.969,0.914}
\def\BibTeX{{\rm B\kern-.05em{\sc i\kern-.025em b}\kern-.08em
    T\kern-.1667em\lower.7ex\hbox{E}\kern-.125emX}}
\begin{document}
\title{Denoising Plane Wave Ultrasound Images Using  Diffusion Probabilistic Models}
\author{Hojat Asgariandehkordi, Sobhan Goudarzi*, Mostafa Sharifzadeh*, Adrian Basarab, \IEEEmembership{Senior Member, IEEE} and Hassan Rivaz, \IEEEmembership{Senior Member, IEEE}
\thanks{This work was supported by the Natural Sciences and Engineering
Research Council of Canada (NSERC).\\
(Corresponding author: Hojat Asgariandehkordi.)
Hojat Asgariandehkordi, Mostafa Sharifzadeh Hassan Rivaz are with the Department
of Electrical and Computer Engineering, Concordia University, Montreal, QC H3G 1M8, Canada (e-mail: Hojat.Asgariandehkordi@mail.concordia.ca; mostafa.sharifzadeh@mail.concordia.ca; hrivaz@ece.concordia.ca.}
\thanks{Sobhan Goudarzi is with Physical Science Platform, Sunnybrook Research Institute, Toronto, ON M4N 3M5, Canada (e-mail: sobhan.goudarzi@sri.utoronto.ca.}
\thanks{Adrian Basarab is with the INSA-Lyon, UCBL, CNRS, Inserm,
CREATIS UMR 5220, Université de Lyon, U1206 Villeurbanne, France
(e-mail: adrian.basarab@irit.fr).}
\thanks{*The authors contributed equally to this work.}}

\IEEEtitleabstractindextext{%
\fcolorbox{abstractbg}{abstractbg}{%
\begin{minipage}{\textwidth}\rightskip2em\leftskip\rightskip\bigskip
\begin{wrapfigure}[15]{r}{3.8in}%
\hspace{-4.1pc}\includegraphics[width=0.55\textwidth]{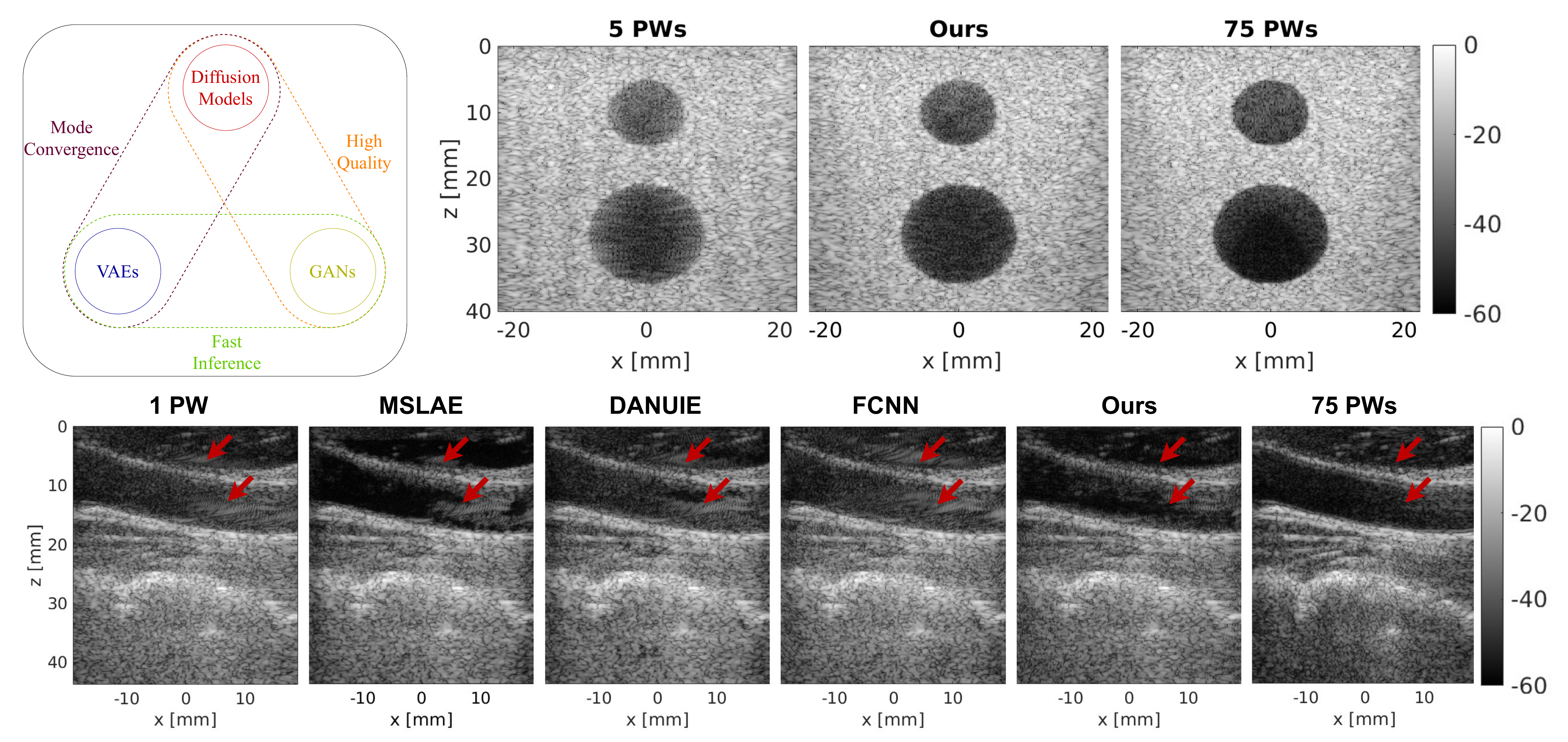}
\end{wrapfigure}%
\begin{abstract}
Ultrasound plane wave imaging is a cutting-edge technique that enables high frame-rate imaging. However, one challenge associated with high frame-rate ultrasound imaging is the high noise associated with them, hindering their wider adoption. Therefore, the development of a denoising method becomes imperative to augment the quality of plane wave images. Drawing inspiration from Denoising Diffusion Probabilistic Models (DDPMs), our proposed solution aims to enhance plane wave image quality. Specifically, the method considers the distinction between low-angle and high-angle compounding plane waves as noise and effectively eliminates it by adapting a DDPM to beamformed radiofrequency (RF) data. The method underwent training using only 400 simulated images. In addition, our approach employs natural image segmentation masks as intensity maps for the generated images, resulting in accurate denoising for various anatomy shapes. The proposed method was assessed across simulation, phantom, and \textit{in vivo} images. The results of the evaluations indicate that our approach not only enhances image quality on simulated data but also demonstrates effectiveness on phantom and \textit{in vivo} data in terms of image quality. Comparative analysis with other methods underscores the superiority of our proposed method across various evaluation metrics. The source code and trained model will be released along with the dataset at: \url{http://code.sonography.ai}
\end{abstract}

\bigskip
\end{minipage}}}
\maketitle

\begin{table*}[!t]
\arrayrulecolor{subsectioncolor}
\setlength{\arrayrulewidth}{1pt}
{\sffamily\bfseries\begin{tabular}{lp{6.75in}}\hline
\rowcolor{abstractbg}\multicolumn{2}{l}{\color{subsectioncolor}{\itshape
Highlights}{\Huge\strut}}\\
\rowcolor{abstractbg}$\bullet$ & We have tailored denoising diffusion models for Radio Frequency (RF) beamformed ultrasound data in plane wave imaging.\\

\rowcolor{abstractbg}$\bullet${\large\strut} & We have shown that the diffusion model can improve RF beamformed data corresponding to single plane wave while preserving the speckle pattern. \\
\rowcolor{abstractbg}$\bullet${\large\strut} & Denoising diffusion models have shown strong performance in denoising natural images, and our work shows their potential in plane wave ultrasound imaging.\\[2em]\hline
\end{tabular}}
\setlength{\arrayrulewidth}{0.4pt}
\arrayrulecolor{black}
\end{table*}

\section{Introduction}
\IEEEPARstart{U}{ltrasound} imaging, a non-invasive and cost-effective medical diagnostic tool, is crucial in modern healthcare by providing real-time visualization of internal structures and abnormalities. Nevertheless, despite its many advantages, the presence of noise can hinder broader use of ultrasound imaging because it can restrict the interpretability and diagnosis precision of the medical images~\cite{10423849}. The development of robust post-processing techniques \cite{QI2021106373}\cite{ESLAMI2022106808} for ultrasound images has, therefore, become imperative to enhance image quality and facilitate more accurate diagnoses.
To address the mentioned challenges, classical techniques have delved into realms such as image filtering, speckle reduction algorithms, and contrast enhancement methods \cite{O'Donnell}. These endeavors, while being impactful, often encounter complexities inherent to ultrasound data, which necessitates the development of more robust methods which have adaptive approaches~\cite{8808885}.

Recently, deep learning has come to the world as a game changer in the domain of medical imaging, and ultrasound image analysis is not an exception. Deep learning has shown remarkable potential in dealing with the formidable challenges of ultrasound imaging. These models, harnessed by their ability to leverage large-scale datasets and discern intricate patterns within ultrasound data, have the power to significantly reduce noise, enhance image contrast, and amplify image fidelity. Recent studies underscore their prowess in tasks such as image denoising, image segmentation, and pathology detection \cite{Shengfeng}. For instance, U-Net \cite{ronneberger2015u} stands as a landmark in medical image segmentation and denoising, spanning several applications in ultrasound imaging \cite{CHEN2023109728}. Furthermore, the work by Kaur \textit{et al}. \cite{Kaur2023} underlines the promising trajectory of deep learning techniques in denoising ultrasound images.

Recently, Denoising Diffusion Probabilistic Models (DDPM) \cite{ho2020denoising} have shown a remarkable performance in image denoising. DDPMs are based on diffusion models, which are a part of larger generative models.
Among the pioneers of generative modeling, Generative Adversarial Networks (GANs) \cite{goodfellow2014}, Variational Autoencoders (VAE) \cite{kingma2013}, and diffusion models \cite{ho2020denoising} stand out as versatile frameworks, each offering a unique perspective on the generation of complex data.

\begin{figure}[h]
  \centering
  \includegraphics[width=0.4\textwidth]{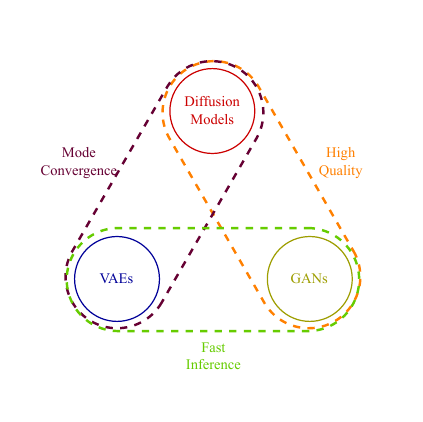}
  \caption{Advantages and disadvantages of the generative models. Diffusion models generate high-quality and diverse results, albeit at a greater computational complexity. }
  \label{fig1}
\end{figure}

To gain a panoramic view of the three mentioned methods, a pictorial overview of the advantages and disadvantages associated with each method \cite{Zhisheng} is presented in Fig.~\ref{fig1}. As illustrated, GANs excel in high-quality sample generation and rapid sampling but face challenges in achieving mode coverage diversity. On the other hand, VAEs possess the advantage of fast inference coupled with mode coverage diversity, although they may not be as effective in generating high-quality samples. Lastly, diffusion models are renowned for their excellence in high-quality sample generation and mode coverage diversity but may lag in terms of sampling time. 

Generative models, when repurposed as denoisers, utilize their learned knowledge of data distributions to remove noise from input data while preserving essential features. This denoising capability makes them valuable tools for enhancing the quality of noisy images, audio, or data. DDPMs stand out among generative models for their ability to reduce noise progressively and iteratively \cite{ho2020denoising}.

Diffusion denoising probabilistic model (DDPM) was the first version of the diffusion generative models, which triggered many other innovations that improved both sampling time and image quality. Those attempts resulted in improved diffusion models~\cite{nichol2021improved},~\cite{karras2022elucidating}, and ``come closer diffuse faster''~\cite{chung2022come}, which decrease the sampling time. Fundamental diffusion models were used to transfer between the normal noise space and the objective training space. However, the most recent diffusion models like ~\cite{liu2022flow,lee2023minimizing,song2020denoising,NEURIPS2023_80fe51a7} have shown the feasibility of moving directly between two arbitrary spaces which are not necessarily normal noise. Such ideas motivated us to use diffusion models to move from the low-quality ultrasound plane waves (PWs) toward the high-quality ones.\\ 
Diffusion models, apart from their applications on natural images, can be leveraged for medical image applications. Medical image segmentation~\cite{dorjsembe2023conditional,10172039}, image inpainting~\cite{durrer2024denoising}, and super-resolution~\cite{han2024arbitrary} are some examples. In ultrasound imaging, recent works have tackled super-resolution~\cite{38636526}, ultrasound dehazing~\cite{10423849}, and ultrasound image reconstruction~\cite{zhang2023ultrasound}.

In this study, we leverage the iterative denoising capabilities of diffusion models to improve the quality of ultrasound images. While we acknowledge in this section that DDPMs may not excel in rapid sampling, it is essential to emphasize that we utilize these models specifically for denoising, not sample generation. As a result, the number of iterations employed for denoising is significantly less than the time typically required when using DDPMs for sample generation. More specifically, the proposed method uses the differences between low-compounded PWs (representing low-quality images) and high-compounded PWs (representing high-quality images) as the perturbation factor instead of the noise (in DDPM) in the forward process. In the reverse process, our method is initialized by a low-quality image (instead of a Normal noise image). This conditioning helps us to reduce the required number of reverse steps, leading to less computations.

A preliminary version of this paper was recently published~\cite{10306544}. This initial publication proposed a method for eliminating Normal noise in ultrasound images, where the input and output of the model were B-mode PW images. Herein, we present the methodology to denoise the RF beamformed PW data, wherein our input and output to the model are RF beamformed images. We also present the method in substantially more depth and extend the validation scope significantly, incorporating evaluations using both real phantom and \textit{in vivo} data. The main contributions of this paper are as follows:
\begin{itemize}
    \item Tailored DDPM for ultrasound image denoising. The training of the DDPM is done on simulated data to learn the unique challenges associated with denoising radiofrequency (RF) data. The inherent strength of the DDPM lies in its fundamental characteristic of decomposing the noise reduction task into a series of incremental steps. 
    \item We show that our network, trained exclusively on simulated data, exhibits strong performance when applied to both phantom and \textit{in vivo} datasets during the testing phase.
    \item Our simulation involves using natural image segmentation masks as echogenicity maps. Consequently, the proposed method can generalize the denoising task to various target shapes, similar to~\cite{10584319} in an aberration correction task.
\end{itemize}

\section{Related work}
Considering the main idea of the proposed method and our emphasis on ultrasound image denoising, in this section, we specifically concentrate on denoising methods. Subsequently, we will delve into the examination of generative models and their application in the context of image denoising. 

\subsection{Ultrasound image denoising methods}
Several elements contribute to the degradation of ultrasound image quality: Gaussian electronic noise present in the RF data, and a range of acoustic noise sources, including reverberation and multiple scattering phenomena. These factors collectively lead to large diffuse reverberations and clutter, adversely affecting image clarity. In this paper, we exploit advances in the field of denoising to improve the quality of ultrasound images. It is important to emphasize that our goal is not to despeckle the ultrasound image, as ultrasound speckle is not simply noise; it often holds essential information pertinent to diagnosis and image-guided interventions. While despeckling might aid tasks like image segmentation, it can adversely affect radiologists and certain subsequent tasks, such as quantitative ultrasound. Instead, our aim is to concentrate on denoising the images, ensuring the preservation of the speckle pattern.

Before deep learning, adaptive filtering methods utilized filters such as the Wiener filters, and adjusted their parameters based on local image characteristics, smoothing out noise while preserving essential anatomical structures \cite{Li-RUi}. Wavelet-based denoising, another prevalent classical method, decomposed ultrasound images into different frequency components, enabling selective noise reduction while retaining crucial image features \cite{Yadav}. The adaptability to target noise at specific scales has made wavelet-based denoising a valuable asset in ultrasound image processing.
Anisotropic diffusion filters were also employed to enhance ultrasound images by selectively permitting diffusion along image edges while preserving the edges themselves \cite{Yadav}. This approach maintained structural details in the image while reducing noise. Furthermore, the non-local means denoising algorithm found its application in ultrasound image denoising \cite{Buades}. Leveraging the redundancy in ultrasound images, the method averaged similar pixels, effectively reducing noise without compromising vital diagnostic information. While classical methods typically suffer from high computational complexity and produce blurry outputs, deep learning-based methods exhibit low computational complexity in the inference phase and generate denoised images of higher quality.

Recently, a number of methods based on deep learning have been developed to improve the quality of ultrasound images. For instance, Perdios \textit{et al.} \cite{Perdios} trained an adapted version of U-Net on simulated data and showed that their network also performs well on \textit{in vivo} data. Gupta \textit{et al.} \cite{Gupta} looked at image denoising as an inverse problem and solved it using a deep learning model. A network called Mimicknet \cite{Huang} was proposed to improve the image quality in post-beamformed data. Zhang et al.~\cite{ZHANG2021102018} aimed to reconstruct high-quality ultrasound PWs from raw channel data by training a self-supervised trained network in an inverse problem approach. Van Sloun et al.~\cite{8808885} have explored deep learning methods that have been developed for adaptive beam-forming, spectral Doppler, clutter suppression, and super-resolution.

Super-resolution is another post-processing technique commonly employed to transform a low-quality image into a higher-quality one. This technique has recently garnered significant attention from researchers specializing in medical image analysis. Khakzad \textit{et al}. \cite{Gharamaleki} designed a network made of an encoder followed by a transformer-based decoder for probable localization. In another similar work\cite{van-Sloun}, Deep-ULM was suggested for vascular ultrasound imaging using super-resolution.

In comparison to natural images, ultrasound data exhibit challenges arising from both high frequency and wide dynamic ranges~\cite{8990076}. These characteristics, irrespective of the specific application, pose significant obstacles to the effective utilization of Convolutional Neural Networks (CNNs) for ultrasound image enhancement. Various studies have attempted to address these challenges by proposing pre-processing techniques, architectures, and loss functions. For instance, it has been demonstrated that individually standardizing each RF ultrasound image in the dataset enhances the performance of CNNs by mitigating the detrimental effects of high dynamic range and utilizing the data more efficiently \cite{10307302}. To address both the high dynamic range and the oscillating properties of RF ultrasound images, mean signed logarithmic absolute error (MSLAE) was introduced as the loss function for training a residual-based CNN with multi-scale and multi-channel filtering properties \cite{Perdios2}. For efficient recovery of high-frequency contents, the wavelet transform was employed in a multi-resolution architecture \cite{GOUDARZI}. To mitigate the risk of getting trapped in local minima during the initial stages of optimization on fluctuating RF ultrasound images, an adaptive mixed loss function was proposed that gradually transitions from B-mode loss to RF loss as the training progresses \cite{10584319}.

One of the major concerns when using CNNs for ultrasound image denoising pertains to the network's ability to retain informative features while removing noise. To address this concern, different attention mechanisms were utilized to guide feature extraction in different layers of CNNs. Dong \textit{et al.} \cite{Dong} employed a multi-stage CNN, introducing feature masking to identify the most beneficial features. Residual attention was another attention mechanism used for more efficient feature selection in \cite{SMA}. In \cite{lan}, image features were effectively extracted, benefitting from a lightweight mixed-attention block to surpass the noise during the encoding. More specifically, this method employs separation and re-fusion techniques for channel-spatial attention. 

Although conventional deep learning-based models have shown a great impact on ultrasound image denoising, challenges such as introducing artifacts to the resulting images and losing image details still exist. In contrast, due to its iterative denoising approach, DDPM effectively reduces noise while preserving crucial image details, offering a superior solution for enhanced image quality. 

\subsection{Generative models}
Numerous valuable efforts have been made to develop generative models for the purpose of denoising and image enhancement. MedGAN\cite{armanious2020medgan}  was designed for medical image-to-image translation, which brought two main contributions. First, a discriminator was trained to play the rule of feature extractor by measuring the distance between the output and the desired one. Second, the textures and fine structures of the desired target images are matched to the translated images using style-transfer losses. Chung et al.~\cite{CHUNG2022102479} have proposed a score-based diffusion model to reconstruct high-quality MRI images from a low-quality image. Their method process the real and imaginary data separately and finally adds them together. In contrast to our method, which is initialized by one PW in the reverse process, their method starts from pure noise and passes many steps to reach a high-quality image in the reverse process. Goudarzi \textit{et al}. \cite{Goudarzi2} proposed a generative adversarial network to recover a high-quality image from a single-focus image, which resembles a multi-focus ultrasound image in terms of quality. UltraGAN\cite{Escobar}  also applied GANs to improve ultrasound echocardiography.

Apart from the GANs, there are some works that adapted diffusion models on ultrasound data. In \cite{10423849} diffusion model was employed for ultrasound cardiography dehazing. Two diffusion models were trained to generate a patch-wise clean image and haze separately using a hazy image as a condition. Domingues \textit{et al}. ~\cite{dominguez2024diffusion} introduced ultrasound physics to a diffusion model for ultrasound image generation. They modified the noise scheduler based on attenuation maps to include the attenuation in their synthetic data. In Zhang \textit{et al.}~\cite{zhang2023ultrasound}, a reconstruction method was proposed based on the diffusion models to reconstruct B-mode images from raw channel data. Their main desire was to eliminate the effects of the normal noise that exists in the raw channel data, which appears as artifacts in the beamformed data. In Li  \textit{et al.}~\cite{li2023single} and Lan  \textit{et al.}~\cite{lan2023fast}, the authors have proposed a score-based diffusion model which trained on B-mode 75 compounding PW images to enhance ultrasound image quality from one PW to 75 compoundings. In a recent work \cite{mishra2023dual}, the authors propose a new method to distinguish 13 anatomies in fetal ultrasound videos using a dual-conditioned diffusion model.

The rest of the paper is designed as follows. The proposed method is described in Section \ref{Method}. Then, the experimental results, datasets, evaluation metrics, and training strategy are detailed in Section \ref{Experiments}. Discussion is in section \ref{Discussion}, and the paper is concluded in Section \ref{Conclusion}.

\subsection{Background}
Denoising diffusion models are composed of two procedures: the forward process and the reverse process. In the forward process, an image is perturbed by normal noise during several steps. In the next step, a network learns how to approximate the noise in each step. In the reverse process, starting from pure noise, the model (using an iterative process) reconstructs a sample using the training distribution based on the learned priors by the network.
 Here are more details about the procedures (from~\cite{ho2020denoising}):\\
\subsubsection{Forward process}
Given an initial image $x_{0}$, the output of each step (a Markov chain) can be defined as follows:
\begin{equation}
x_{t} = \sqrt{1 - \beta_t} x_{t-1} + \sqrt{\beta_t} \epsilon, \quad \epsilon \sim \mathcal{N}(0, I),
\end{equation}
where $x_t$ is the data at time step $t$, $\beta_t$ is a variance schedule, $\epsilon$ is noise sampled from a normal distribution $\mathcal{N}(0, I)$. In that case, $q(x_{t}|x_{t-1})$ has a normal distribution:
\begin{align}
q(x_{t} | x_{t-1}) = \mathcal{N}(x_{t-1},\mu(x_{t-1},t), \sigma_t).
\label{eq111}
\end{align}
$x_t$ can be directly linked to $x_0$ by:
\begin{equation}
x_t = \sqrt{1 - \overline{B}_t} x_0 + \sqrt{\overline{B}_t} \epsilon,  where, \overline{B}_t=\sum_{s=1}^{s=t} B_{s}
\end{equation}
\subsubsection{Reverse process}
In the reverse process, the objective is to transition from a noisy (low-quality) image at step $T$ and enhance the image quality step by step to approximate the original high-quality image $x_0$ closely. This entails maximizing the likelihood of the predicted image $P_{\theta}$, where $\theta$ denotes the prediction parameters, or minimizing the negative log-likelihood of the prediction as follows:
\begin{align}
P_{\theta}(x_0) = \int p_{\theta}(x_{0:T}) \, dx_{0:T},
\label{eq1}
\end{align}
\begin{align}
L_{\theta}(x_0)= -\log(P_{\theta}(x_0)),
\label{eq2}
\end{align}
where
\begin{align}
p_{\theta}(x_{0:T})=\prod_{t=1}^{T} p_{\theta}(x_{t-1}|x_{t}).
\label{eq4}
\end{align}
Note that $p_{\theta}$ is the predicted noise distribution at each step of the reverse process, and $P_{\theta}$ indicates the PDF of the predicted $x_0$. Since the predicted $x_0$ is conditioned on several consecutive steps, to calculate $P_{\theta}$ in integral form, we need to integrate over a very high dimensional (pixel) space for continuous values over $T$ time steps. As such, this integral is not tractable. To solve the problem stemming from variational lower bound \cite{sohl2015deep}, one can write:
\begin{align}
-\log(P_{\theta}(x_0)) &= -\log\left(\int p_{\theta}(x_{0:T}) \, dx_{0:T}\right) \nonumber \\
&= -\log\left(\int p_{\theta}(x_{0:T}) \frac{q(x_{0:T}|x_{0})}{q(x_{0:T}|x_{0})} \, dx_{0:T}\right) \nonumber \\
&= -\log\left(E_q\left[\frac{p_{\theta}(x_{0:T})}{q(x_{0:T}|x_0)}\right]\right), 
\label{eq5551}
\end{align}
where $E_q$ is the marginal expected value of $q$. Based on Jensen's inequality, one can write:
\begin{align}
-log(E_q[\frac{p_{\theta}(x_{0:T})}{q(x_{0:T}|x_0)}])\leq E_q[-log(\frac{p_{\theta}(x_{0:T})}{q(x_{0:T}|x_0)})].
\label{eq666}
\end{align}
Incorporating (\ref{eq5551}) and (\ref{eq666}), we will have:
\begin{align}
-log(P_{\theta}(x_0))\leq E_q[-log(\frac{p_{\theta}(x_{0:T})}{q(x_{0:T}|x_0)})].
\label{eq7}
\end{align}
Therefore,
\begin{align}
-log(P_{\theta}(x_0))\leq E_q[-\log(p_{\theta}(x_T)-\sum_{t<T}\log(\frac{p_{\theta}(x_{t-1}|x_{t})}{q(x_{t}|x_{t-1})}))].
\label{eq8}
\end{align}
Note that $p_{\theta}(x_T)$, the start point of the reverse process, is not predicted by the network's parameters. Hence, to minimize the right side of the above equation, the following term should be minimized.
\begin{align}
\sum_{t<T}\log(\frac{p_{\theta}(x_{t-1}|x_{t})}{q(x_{t}|x_{t-1})}).
\label{eq9}
\end{align}
Using Bayes' rules, \ref{eq9} becomes:
\begin{align}
\sum_{t<T}\log(\frac{p_{\theta}(x_{t-1}|x_{t},x_0)q(x_t|x_0)}{q(x_{t-1}|x_{t})q(x_{t-1}|x_{0})}).
\label{eq10}
\end{align}
After expanding the sum and doing some simplifications, it can be proven that the following equation needs to be satisfied to minimize that term.
\begin{align}
p_{\theta}(x_{t-1}|x_{t})={q(x_{t-1}|x_{t},x_0)}, 0<t<T
\label{eq444}
\end{align}
Considering (\ref{eq111}), we have: 
\begin{align}
q(x_{t-1} | x_{t},x_0) = \mathcal{N}(x_{t-1},\tilde{\mu}(x_t,t), \tilde{\sigma(t)}).
\label{eq555}
\end{align}
Therefore, based on~\ref{eq444}, 
\begin{align}
p_{\theta}(x_{t-1} | x_{t}) = \mathcal{N}(x_{t-1},\mu_{\theta}(x_t,t), \sigma(t)).
\label{eq13}
\end{align}
To satisfy that, a network should be trained to:
\begin{align}
\theta = \arg\min_{\theta} \left\{l_1(\mu_{\theta}(x_t,t),\tilde{\mu}(x_t,t)) \right\},
\end{align}
Because $\mu$ is a function of the additive noise in each step ($\epsilon$), the network should learn the additive noise in the forward process to be able to approximate $x_{t-1}$ in each reverse step. Hence:
\begin{align}
\theta = \arg\min_{\theta} \left\{l_1(\epsilon_{\theta}(x_t,t),\epsilon(x_t,t)) \right\},
\end{align}
 where, $\epsilon_{\theta}$ points to the approximated noise by the network.\\
 \\
 
\section{Methods}\label{Method}
\begin{figure}[h]
  \centering
  \includegraphics[width=0.53\textwidth]{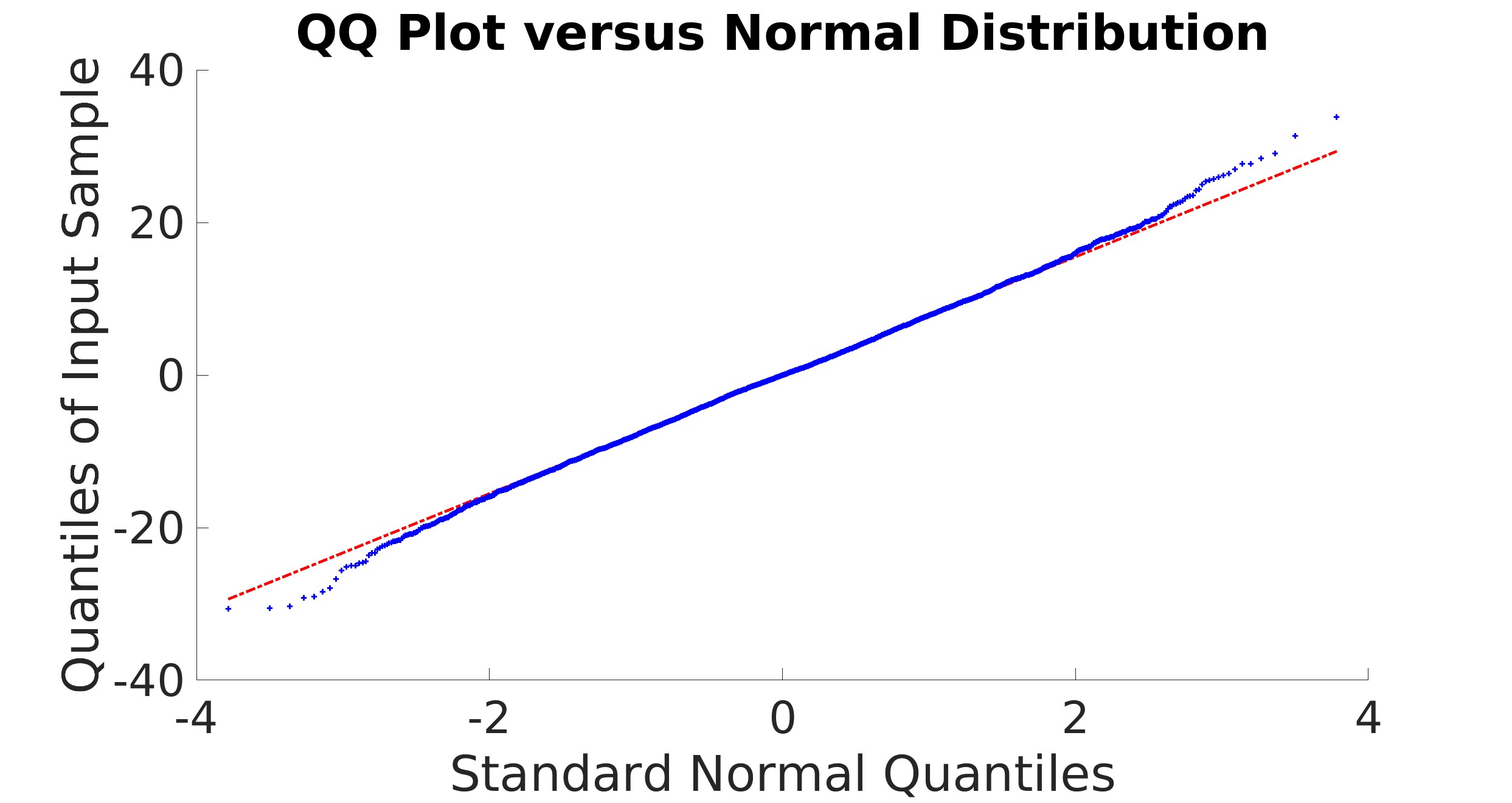}
  \caption{The QQ Plot corresponding to the difference between RF data reconstructed with one and 75 angles. A linear trend highlights a Gaussian distribution. The difference follows a Gaussian distribution in the middle, with deviations from this distribution in the first and last quantiles.}
  \label{fig222}
\end{figure}

This paper presents an iterative model inspired by DDPMs to enhance a low-quality beamformed RF image to a high-quality beamformed RF image. Specifically, our primary objective is to enhance the quality of a PW image initially constructed with a restricted number of compounded angles to closely resemble the quality of an image constructed using a larger number of angles. In other words, we consider the low-compounded ultrasound images (low-quality) in $\pi_1$ space and their corresponding high-compounded counterparts (high-quality) in $\pi_0$. Therefore, our problem can be defined as a transition from $\pi_1$ to $\pi_0$. \\
In similar diffusion-based denoising methods~\cite{10423849,chung2022come}, a diffusion model is trained on high-quality image space to generate high-quality images (corresponding to $t=0$) starting from Normal noise (corresponding to $t=T$). Then, in the inference procedure, a low-quality image passes through a number of forward process steps to the extent that the resulting noisy image becomes close to the noise in the training phase (corresponding to $t=T$). Finally, in the reverse process, the trained network generates a high-quality image starting from the noisy image corresponding to $t=T$. This approach includes many steps in the reverse processes, which creates very high-quality results but requires additional guidance to ensure the reliability of the output. As an example, Li~\textit{et al.}~\cite{li2023single} also adapted this approach and introduced data consistency to reduce the error that can be compounded in several diffusion steps. In this work, similar to~\cite{liu2022flow, lee2023minimizing, song2020denoising,NEURIPS2023_80fe51a7}, the idea is to train a diffusion model to move directly from $x_1$ in space $\pi_1$ toward $x_0$ in space $\pi_0$. This approach dramatically reduces the number of required sampling steps because a single-PW image is not very different from a compounded image.\\
Based on our particular objective, we examined the disparity between low-compounded ultrasound images (low-quality from $\pi_1$) and their corresponding high-compounded counterparts (high-quality from $\pi_0$). The analysis indicates that the distribution of this difference can be close to a Gaussian distribution. To visually depict this resemblance, a Quantile-Quantile plot (QQ-plot) has been generated, illustrating the difference between beamformed RF data corresponding to one and 75 compounding PWs patches in Fig~\ref{fig222}. Having a close look at the plot, there are some unfitted regions between the blue points and the red line (Gaussian distribution), which means the mentioned difference is not necessarily Gaussian in some regions.

Although the vanilla diffusion models used normal  Gaussian noise in their forward and reverse processes, here, stemming from~\cite{lee2023minimizing,song2020denoising,NEURIPS2023_80fe51a7}, we define a forward and a reverse process to directly transfer between $\pi_0$ and $\pi_1$ without requiring to go to Normal noise distribution in between. It means the forward and reverse processes are not merely dependent on Gaussian noise.

\subsection{Forward process}
\begin{figure*}[t]
  \centering
  \includegraphics[width=1\textwidth]{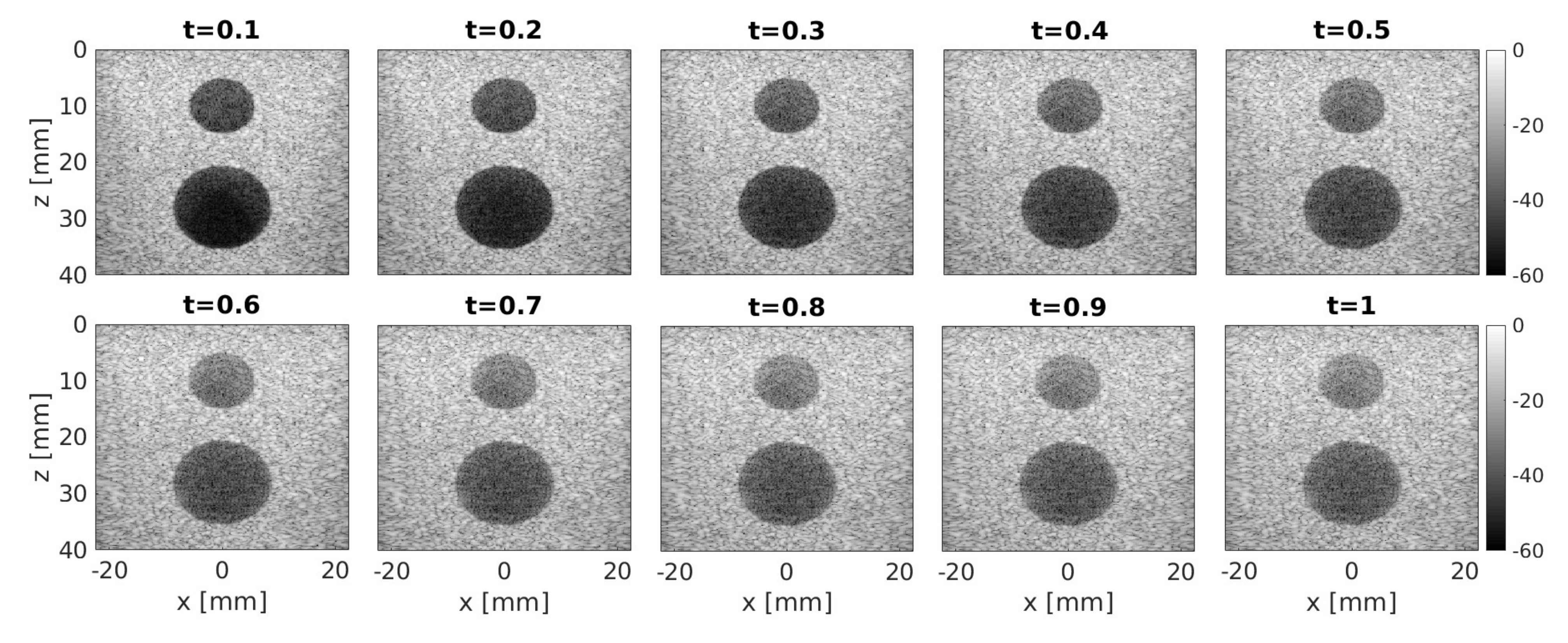}
  \caption{An example of the forward diffusion process in 10 steps. The training occurs in the reverse process, while no learning takes place during the forward process.}
  \label{fig333}
\end{figure*}

Having observations like $X_0\sim\pi_0$ and $X_1\sim\pi_1$, we can construct a forward process based on ordinary differential equations to gradually move from $\pi_0$ to $\pi_1$ following an interpolation procedure:
\begin{align}
x_t = (1-t)X_0 + t X_1,
\label{eq1411}
\end{align}
 where $t \in [0,1]$. After a minor replacement:
\begin{align}
x_t = X_0 + t (X_1-X_0),
\label{eq1511}
\end{align}
 \begin{align}
\Delta x_t =(X_1-X_0) \Delta t, 
\label{eq1611}
\end{align}
 \begin{align}
x_{t+\Delta t} = x_t+\Delta x_t=x_t+v(x_t,t)\Delta t .
\label{eq1711}
\end{align}
Hence, $(X_1-X_0) $ can be considered as a flow velocity $v(x_t,t)=X_1-X_0$, which determines the changes in each time step (in the forward process).  Figure~\ref{fig333} illustrates an example of the described process for $T=10$ ($t \in \{0, 0.1 , 0.2, ... , 1\}$, $\Delta t=0.1$), wherein a high-quality image is gradually affected by $v(x_t,t)$ over ten steps.\\
\subsection{Reverse process}
In the reverse process, starting from an initial point like $x_1$ from $\pi_1$, we want to move toward $x_0$ from $\pi_0$ while following a trajectory that is as close as possible to the reverse of the forward process, which means maximizing $p_{\theta}(x_1)$. Similar to~\ref{eq444} and~\ref{eq555}:
 \begin{align}
q(x_{t-\Delta t} | x_{t},x_0) = q(x_{t}-v(x_{t-\Delta t},t)\Delta t).
\label{eq1811}
\end{align}
 \begin{align}
p_{\theta}(x_{t-\Delta t} | x_{t}) = p(x_{t}-v_{\theta}(x_{t-\Delta t},t)\Delta t).
\label{eq1911}
\end{align}
To make sure that we have a true trajectory in the reverse process, a network should be trained to solve the following optimization problem:
\begin{align}
\theta = \arg\min_{\theta} \left\{l_1(v(x_{t},t), v_{\theta}(x_{t},t)) \right\},
\end{align}
 where $l_1$ denotes the loss function and $\theta$ refers to the network`s parameters.
Finally, the reverse process is done as follows:
 \begin{align}
x_{t-\Delta t}= x_{t}-v_{\theta}(x_{t-\Delta t},t)\Delta t.
\label{eq2011}
\end{align}
It is worth noting that because there is no stochastic factor in the reverse process, and also our network does not have any randomness factor, the reverse process is considered a deterministic process.

\subsection{Architecture}
\begin{figure*}[t]
  \centering
  \includegraphics[width=1\textwidth, height=4in]{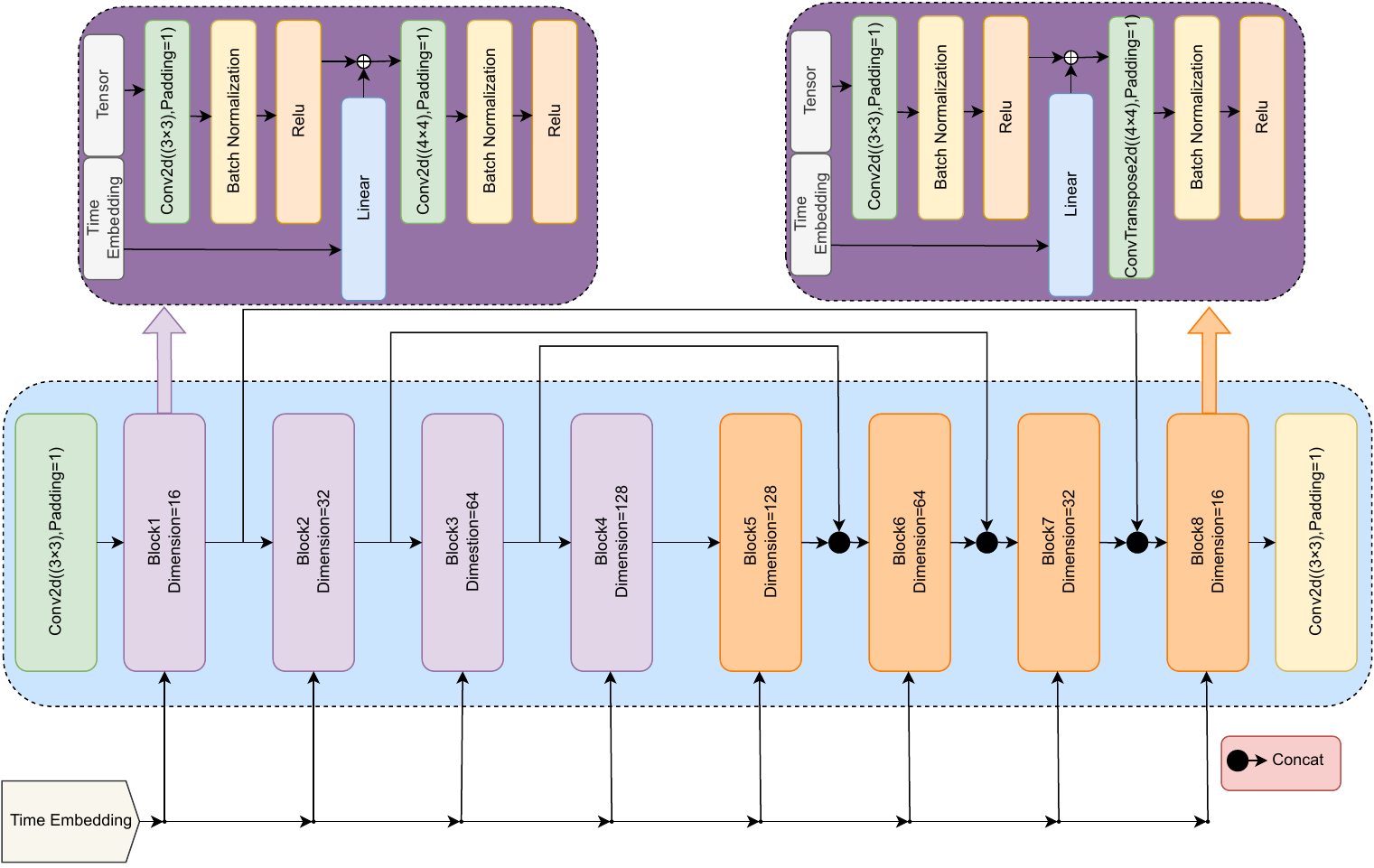}
  \caption{The proposed architecture, which is inspired by the U-Net.}
  \label{fig4}
\end{figure*}

The overarching structure of the employed CNN is illustrated in Fig.~\ref{fig4}, which is comprised of two primary modules called convolutional blocks and time embedding modules.
The time embedding module is devised to allocate a vector to each time step $t$ through Sine and Cosine transforms, serving as informative cues for the network's temporal understanding. These time vectors are subsequently input into the convolutional blocks to provide temporal context. Within the convolutional blocks, the data stream incorporates an input tensor alongside the associated time vectors. Initially, the input tensor undergoes a 3$\times$3 convolutional layer, succeeded by batch normalization and ReLU nonlinearity layers. Simultaneously, the input time vector is processed through a linear layer to align with the features extracted from the convolutional layers, facilitating their summation in the subsequent step. Subsequently, the resulting tensors traverse a 4$\times$4 convolutional layer, followed by batch normalization and ReLU activation functions. Note that the provided description pertains to the first four blocks, while for the subsequent blocks, a 4$\times$4 convTranspose is employed instead of the 4$\times$4 convolutional layer. In all convolution layers, both the stride and padding were set to one.

\section{Experiments}\label{Experiments}
This section provides details regarding the evaluation procedure for the proposed method. We begin with an overview of the datasets, followed by an explanation of the data generation sequences and the inclusion of test images. Subsequently, we delve into the evaluation metrics and then elaborate on the training strategy and comparison results.
\subsection{Dataset}
In this study, the proposed network was trained using a simulated dataset. We leveraged both phantom and \textit{in vivo} data for evaluation, supplementing the simulated test data.
\subsubsection{Simulated data}\label{Simulated data}
\begin{figure}[h]
  \centering
  \includegraphics[width=.8\columnwidth]{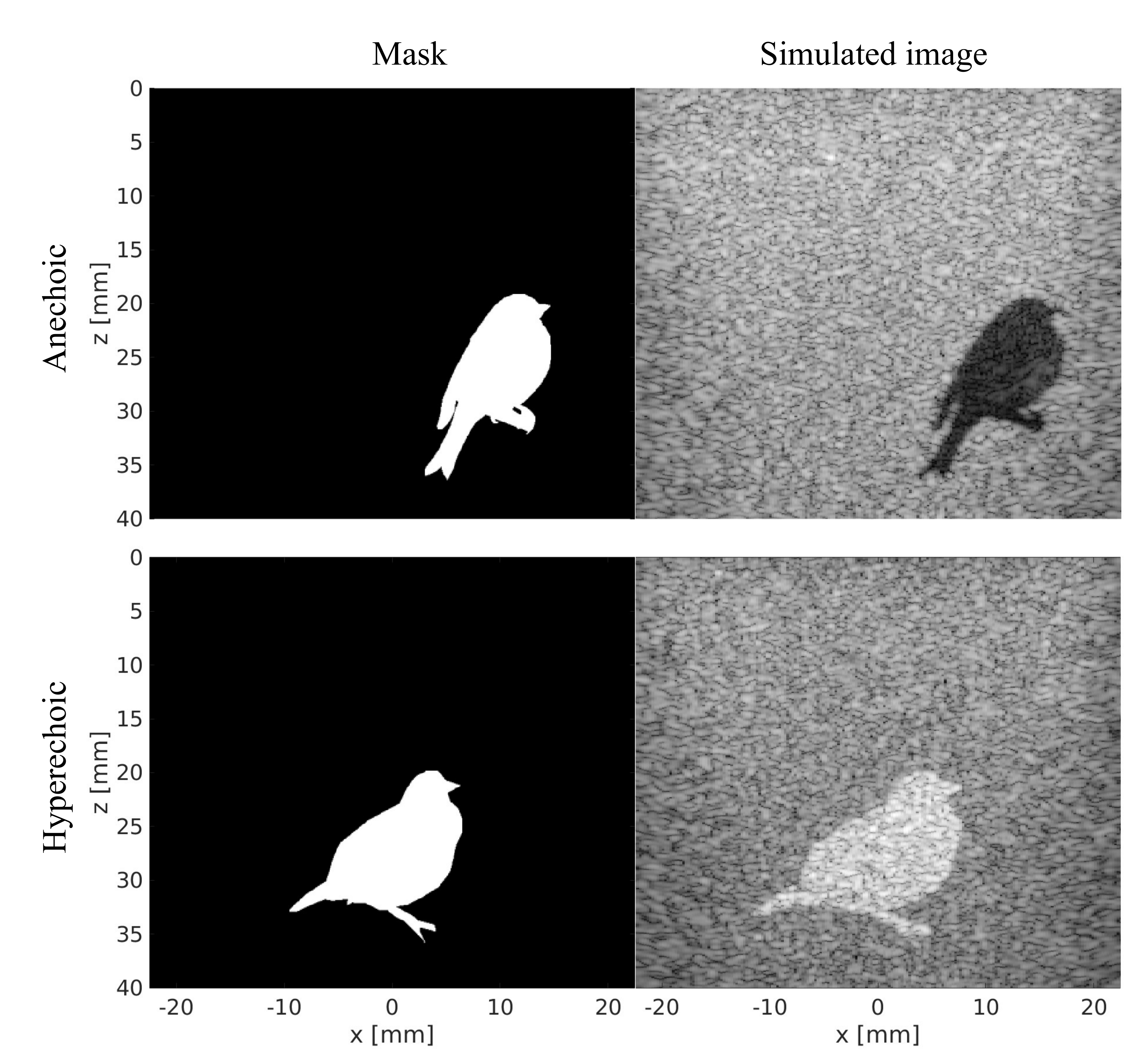} 
  \caption{Samples of mask-simulated images used for the training procedure.}
  \label{fig5}
\end{figure}

The simulated dataset, which consists of 400 images, was simulated using the publicly available Field II package~\cite{jensen1992calculation}~\cite{jensen1997field}. The designed phantom, situated at an axial depth of 10 mm from the transducer's face, measures 45 mm laterally and 40 mm axially and contains a uniform distribution of fully developed speckle patterns, with an average scatterer density of 60 per resolution cell. Two different types of contrast were introduced to the images, including anechoic regions and hyperechoic regions. To create the two echogenicity types and similar to \cite{10584319}, we utilized a public dataset called XPIE \cite{Xia}, which contained segmented natural images. We isolated only the segmentation masks and resized them to match the dimensions of the phantoms. Subsequently, we applied a weight to the amplitude of scatterers within the masks, determined by the echogenicity type of each image. For anechoic regions, the weight was set to zero. For hyperechoic regions ranging from +3 dB to +12 dB, the weight was a uniform random number in the interval [2, 15.8]. The image simulation method outlined in this subsection has the benefit of supplying the network with more extensive features than just including simple cyst-shaped regions. Furthermore, a test image was generated, depicting two cysts with diameters of 10 mm and 15 mm. These cysts were located at central lateral positions, with one at a depth of 10 mm and the other at a depth of 28 mm. The transducer defined for the simulation mirrors the 128-element linear probe L11-5v (Verasonics in Kirkland., WA). The values opted for central and sampling frequencies were 5.208 MHz and 20.832 MHz, respectively. Note that Field II is subject to numerical precision limitations, compelling us to establish the initial sampling frequency at 104.16 MHz and, subsequently, downsample the resulting data by a factor of 5. All images were simulated using a full synthetic aperture scan. Following that, for each image, we synthesized 75 PW scans corresponding to different emission angles. After synthesizing the raw RF data, the delay and sum (DAS) algorithm was applied to construct beamformed RF images with dimensions of 1082$\times$192 in the axial and lateral directions, respectively. Fig.~\ref{fig5} depicts the selected mask and its resulting simulated 75 compounding PW image for two samples in the training set. The compounding angle range started from -16 to 16 degree with the difference of 0.5 degree for each PW.\\

\subsubsection{Phantom and \textit{in vivo} data}
In addition to the evaluation on simulated test images, the performance of the proposed method has been assessed on experimental phantom and \textit{in vivo} images. Both phantom and \textit{in vivo} images are from publicly available PICMUS~\cite{Liebgott} benchmark dataset. The following explains more about each image used in our test phase:
\begin{enumerate}[label={\arabic*},align=left]
  \item Experimental Contrast Phantom: The image was acquired from a CIRS phantom (Model 040GSE), specifically representing areas with anechoic cysts against a speckle background, to evaluate contrast.
  \item \textit{in vivo} image: The image was taken from the longitudinal view of a volunteer's carotid artery.
\end{enumerate}
\subsection{Evaluation metrics}\label{Evaluation metrics}\color{black}
We use key metrics like Contrast to Noise Ratio (CNR) and Generalized Contrast to Noise Ratio (gCNR)\cite{Rodriguez} for quantitative evaluation. In addition, because our goal is enhancing the image quality using a high-quality image as the target, we should investigate how much our method can get the input image closer to the target image. To do so, we measure the similarity between the output images and their corresponding high-quality image utilizing Normalized Root Mean Squared Error (NRMSE).

1) The CNR index is calculated as follows:
     \begin{align}
     \mathrm{CNR} = 20 \log_{10}\left(\frac{\left| \mu_{\text{ROI}} - \mu_B \right|}{\sqrt{{(\sigma^2_{\text{ROI}} + \sigma^2_B)}/2}}\right),
    \label{eq23}
    \end{align}
    where $\mu$ and $\sigma$ point to the mean and standard deviation of 
 both background and region of interest, respectively.
 
 2)  It has been proved that CNR depends on dynamic range alternation~\cite{8918059}, and gCNR was introduced to mitigate this problem. The corresponding formula for gCNR calculation is:

    \begin{align}
    \mathrm{gCNR} = 1 - \int_{-\infty}^{\infty} \min\{p_{\text{ROI}}(x), p_B(x)\} \, dx ,
    \label{eq24}
    \end{align}
where $p_{\text{B}}$ and $p_{\text{ROI}}$ points to the histogram of the background and region of interest, respectively. The gCNR metric quantifies the extent of overlap between the pixel intensity distributions in two regions, considering dynamic range transformations. A higher gCNR value signifies reduced overlap between the distributions, reaching a maximum value of 1 when the two distributions exhibit no overlap.

3) NRMSE: The Euclidean distance, renowned alongside the Mean Squared Error (MSE), stands as one of the most prevalent metrics employed for assessing the similarity between two vectors. However, when evaluating diverse methods across disparate datasets characterized by specific dynamic ranges, computing the general MSE may fail to provide a comprehensive understanding of the enhancement potential. Therefore, normalization is employed to scale distances within the range of [0, 1] across all test sets. For an image denoted as $I$ and a corresponding target $Y$, comprising $L$ elements, NRMSE can be calculated as follows:

   \begin{align}
    \mathrm{RMSE} = \sqrt{\frac{1}{L} \sum_{i=1}^{L} (I_i - Y_i)^2},\
    \label{eq25}
    \end{align}

    \begin{align}
    \mathrm{NRMSE} = \frac{\mathrm{RMSE}}{\mathrm{max}(Y)}.
    \label{eq26}
    \end{align}

\noindent Note that the denominator corresponds to the range of $Y$, as the minimum value of $Y$ is zero.


4) KS test: The Kolmogorov-Smirnov (KS) test in ultrasound, is a statistical method used to quantifies the differences between the cumulative distribution functions (CDFs) of pixel intensity values in the images, providing a measure of how well the processed image retains the characteristics of the original speckle pattern.\\

5) SSIM : The Structural Similarity Index (SSIM) is a metric used to measure the similarity between two images. It takes into account three components of image quality: luminance, contrast, and structure. SSIM compares local patterns of pixel intensities in the images and computes a score between -1 and 1, where 1 indicates perfect similarity. Higher SSIM values suggest that the images are more similar in terms of quality.
{\subsection{Training}
The architecture illustrated in Fig.~\ref{fig4} was employed to predict noise parameters at each time step. During the training phase, following the forward process, a high-quality beamformed image undergoes degradation based on a randomly selected time step noise coefficient $\overline{\beta_t}$, where $1 < t < T$, as described in \cite{ho2020denoising}. Subsequently, the network is exposed to the degraded image as input and its corresponding additive noise as output. This enables the network to learn how to predict the noise parameters without bias towards any specific time step. The training procedure encompassed 350 epochs, employing an Adam optimizer \cite{kingma2014adam} with zero weight decay. The initial learning rate was set to 0.004 and was adjusted by a linear scheduler with a step size of 60 and a gamma value of 0.5 during the training stage. While the designed network is independent of the image size, all simulated images were in the size of 1082$\times$192. Among the 400 simulated images, 20 and 10 percent were randomly selected as validation and test sets, respectively. The remaining images constituted the training set. An $l_1$ norm loss function was utilized to quantify differences between ground truth and predictions during the training phase. All training procedures were conducted on an NVIDIA RTX 4090 GPU with 24 GB RAM.

\subsection{Results}
The proposed method relies on three parameters: $J$ (the number of compoundings in a low-quality image), $K$ (the number of compoundings in a high-quality image), and $T$ (the number of diffusion steps). The results presented in this section are based on the parameters that yielded superior outcomes in the conducted experiments and will be specified during the explanation of each experiment.

\subsubsection{Results on simulated test data}
To evaluate the performance of the proposed method within the same domain as the training dataset, we generated a test image with the same configuration as the training dataset, as detailed in subsection~\ref{Simulated data}. The proposed method was then evaluated using the simulated test image across three distinct scenarios explained in Table~\ref{TABLE1}, each featuring specific low-quality and high-quality pairs. The quantitative results corresponding to each scenario are outlined in Table~\ref{TABLE2}, and the network outputs for each scenario are shown in Fig.~\ref{fig6}.

\begin{table}[h]
\caption{Evaluation with different scenarios. $J$ refers to low compounding and $K$ refers to high compounding}
\centering
\begin{tabular}{|c|c|c|c|}
  \hline
  Scenario & $J$ & $K$ & $T$ (steps) \\
  \hline
  1 & 1 & 75 & 10 \\
  2 & 5 & 75 & 10 \\
  3 & 5 & 15 & 5 \\
  \hline
\end{tabular}
\label{TABLE1}
\end{table}

In Fig.~\ref{fig6}, the first row showcases the results obtained by feeding a single-angle PW image into the network, which was originally trained on ground truth data comprising 75 compoundings. The second row illustrates the output of the proposed method when presented with five-angle compounding PWs. Lastly, the third row demonstrates the enhancement of a five-angle compounding image, even though the target during the training phase was a 15-angle compounding image. As can be observed in Fig.~\ref{fig6}, the proposed method consistently improves image quality in all scenarios. 

Beyond visual comparisons, the proposed method is also quantitatively assessed in each scenario, and results in Table~\ref{TABLE2} confirm the achieved improvement in image quality in terms of all calculated indexes. As discussed in subsection~\ref{Evaluation metrics}, specific regions in the test images are defined as foreground and background for gCNR and CNR calculation. These regions are delineated by green and red circles, respectively, in the last row of Fig.~\ref{fig6}. Notably, the proposed method exhibits striking improvements compared to the input images, particularly evident in the contrast evaluations (gCNR and CNR). Regarding NRMSE, our method effectively bridges the gap between the input images and the target values. These improvements are consistently observed across all scenarios. Hence, we conduct the rest of the experiments with the input images requiring the least number of compoundings (training based on scenario 1). 
  

\begin{figure*}[t]
  \centering
  \includegraphics[width=1\textwidth]{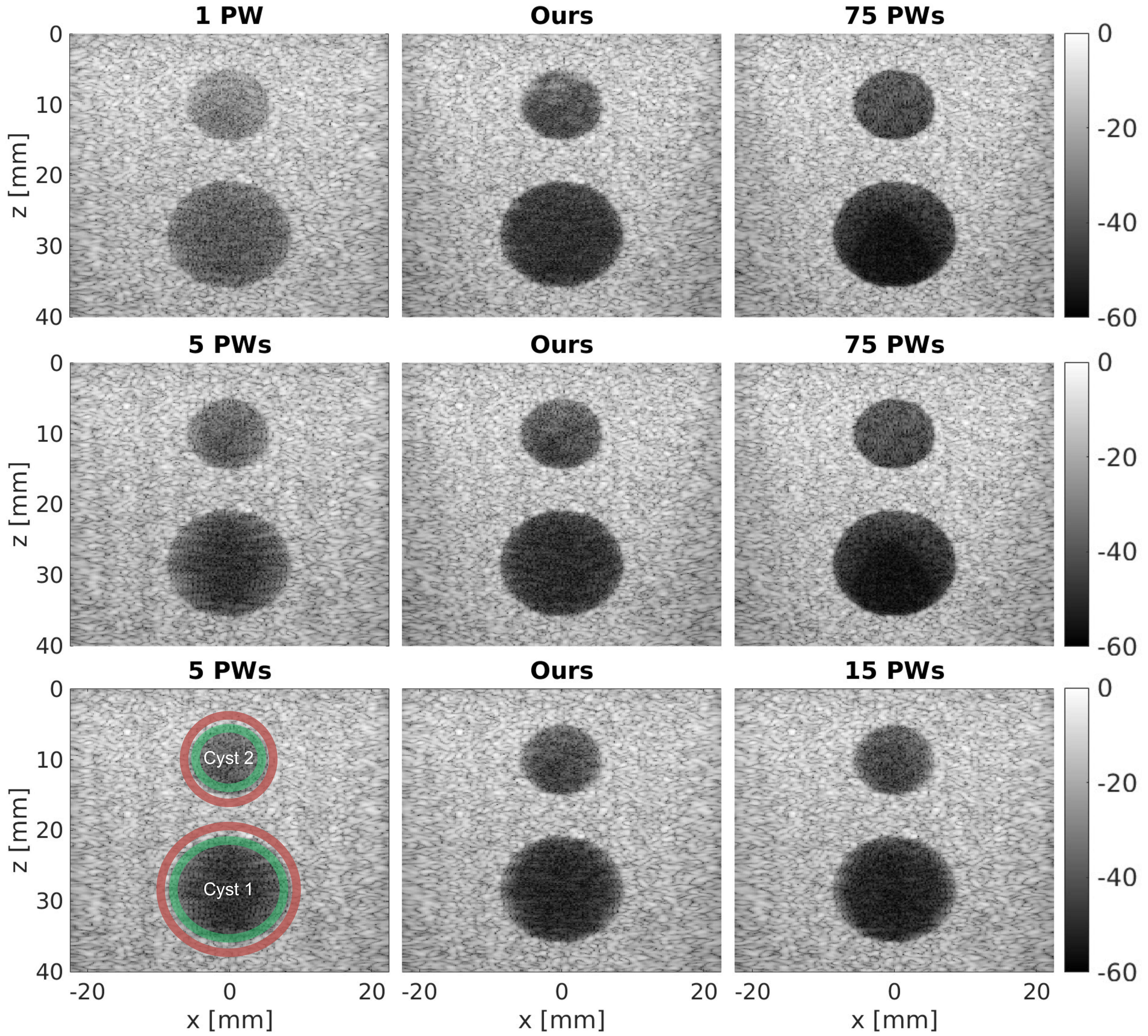}
  \caption{The results on the simulated test image for different scenarios. PW denotes PW. The circular regions in the last row are used for calculating evaluation metrics.}
  \label{fig6}
\end{figure*}

\begin{table}[h]
\caption{Results on the simulated test image.}
\resizebox{\columnwidth}{!}{
    \begin{tabular}{|c|c|c|c|c|}
    
        \hline
        Metric & Scenario & Input Image & Ours & Target image \\
        \hline
        \multirow{3}{*}{NRMSE} & \multirow{1}{*}{1} & \multirow{1}{*}{0.122} & \multirow{1}{*}{0.043} & \multirow{1}{*}{-} \\
        \multirow{3}{*}{(compared to target)}& 2 & 0.067 & 0.033 & - \\
        & 3 & 0.035 & 0.02 & - \\

        \hline
        \multirow{3}{*}{CNR (Cyst 1)} & \multirow{1}{*}{1} & \multirow{1}{*}{7.1 db} & \multirow{1}{*}{10.2 db} & \multirow{1}{*}{11.6 db} \\
        & 2 & 7.3 db & 10.3 db & 11.6 dB \\
        & 3 & 7.3 db & 8.5 db & 8.8 dB\\

        \hline
        \multirow{3}{*}{CNR (Cyst 2)} & \multirow{1}{*}{1} & \multirow{1}{*}{4.6 db} & \multirow{1}{*}{6.4 db} & \multirow{1}{*}{8.6 db} \\
        & 2 & 5.8 db & 8 db &8.6 dB \\
        & 3 & 5.8 db & 6.8 db & 6.6 dB\\

        \hline
        \multirow{3}{*}{gCNR (Cyst 1)} & \multirow{1}{*}{1} & \multirow{1}{*}{0.59} & \multirow{1}{*}{0.73} & \multirow{1}{*}{0.85} \\
        & 2 & 0.63 & 0.71 & 0.85\\
        & 3 & 0.63 & 0.65  & 0.66\\

        \hline
        \multirow{3}{*}{gCNR (Cyst 2)} & \multirow{1}{*}{1} & \multirow{1}{*}{0.76} & \multirow{1}{*}{0.78} & \multirow{1}{*}{0.86} \\
        & 2 & 0.77 & 0.81 &0.86 \\
        & 3 & 0.77 & 0.80 &0.81 \\

        \hline
    \end{tabular}
    }
    \label{TABLE2}
\end{table}

\subsubsection{Results on phantom data}
In this section, we aim to evaluate our method using the CIRS phantom provided in the PICMUS dataset~\cite{Liebgott}. Alongside showcasing our method's results, we also trained the network architectures proposed in \cite{Perdios}, \cite{Perdios2}, and \cite{GOUDARZI}, in a fully supervised fashion, on our training set to enable a comparison with other methods. Fig.~\ref{fig7} visually demonstrates the comparisons on the phantom data. Overall, all methods have succeeded in denoising the input image. However, a closer examination of details, particularly in the magnified cysts, provides more insights into each model's denoising potential. In the cyst patches, it is observed that other methods have partially improved the distinguishability of cysts compared to the input image, yet the boundaries of the cysts are not clearly defined. When it comes to the proposed method, we found that due to the domain shift between our simulation training dataset and experimental test data, high iterations of the reverse process may cause some artifacts in the resulting image. Therefore, we set the number of reverse iterations equal to 3, 5, and 7, respectively. The resulting images across all iterations reveal that cyst areas are more detectable compared to other methods. Notably, in the fifth and seventh iterations, the boundaries of the cysts are even more clear than in the 75 compounded image. In addition, there are distortions in the reconstructed images in cyst 2, which are most pronounced in (b) to (d). These results correspond to MSLAE \cite{Perdios2}, DANUIE \cite{Perdios}, and FCNN \cite{GOUDARZI}, respectively. These distortions demonstrate the extent to which various methods can reduce clutter in the anechoic region.

\begin{figure*}[t]
  \centering
  \includegraphics[width=1\textwidth]{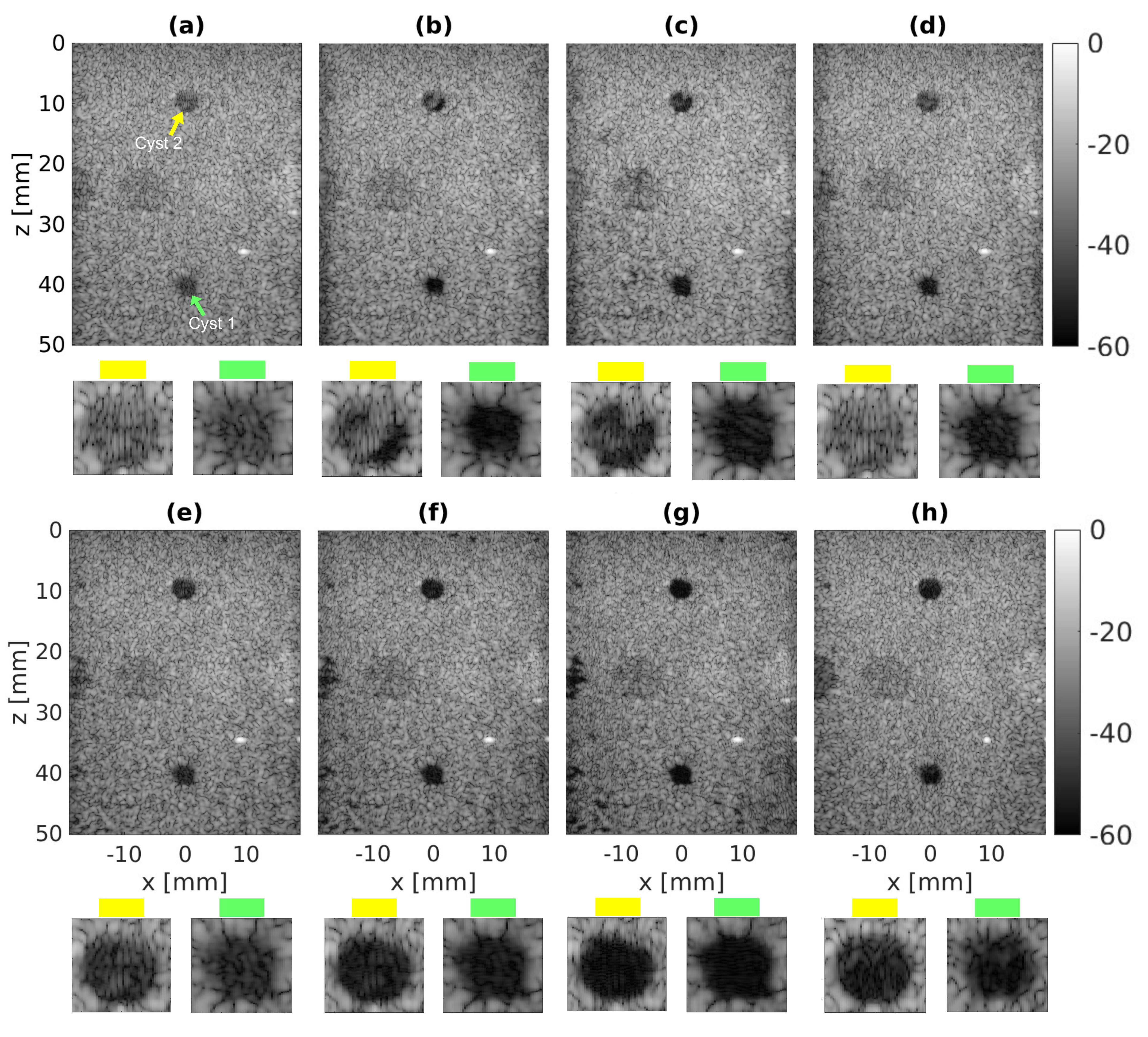}
  \caption{The results for the experimental phantom image. (a) image reconstructed using single PW. (b)  MSLAE\cite{Perdios2} result, (c) DANUIE \cite{Perdios} result, (d) FCNN \cite{GOUDARZI} result, (e) ours with 3 iterations, (f) ours with 5 iterations, (g) ours with 7 iterations (h) image reconstructed using 75 PWs.}
  \label{fig7}
\end{figure*}

The CNR and gCNR values of each cyst in Fig.~\ref{fig7} are listed in Table~\ref{TABLE3} to validate the performance of different methods. The CNR 1 and CNR 2 results represent that our proposed enhancement method outperforms the other methods, achieving the highest CNR values of 6.5 dB for CNR 1 and 8.5 dB for CNR 2, which is higher than the 75 compounding result. Additionally, we evaluated gCNR values, which provide a more comprehensive assessment by considering global image characteristics. Ours (5 iters) consistently yields the highest gCNR values of 0.9560 for gCNR 1 and 0.9629 for gCNR 2, surpassing the other methods. Due to the fact that our method performs better with five iterations as compared to seven iterations, we did not include the corresponding results. In terms of structural similarity, our proposed method surpasses the others with 0.80 and 0.81 SSIM, respectively. As for the speckle test, all of the methods preserve the speckle pattern because they have passed the KS test.\\
Table~\ref{TABLE4} compares the performances in terms of the NRMSE metric. DANUIE\cite{Perdios}, MSLAE\cite{Perdios2}, and FCNN \cite{GOUDARZI} exhibit comparable accuracy with NRMSE values of 0.030, 0.027, and 0.030, respectively, while the proposed method stands out with three iterations, achieving a notably lower NRMSE of 0.0196. Despite a slight increase to 0.0240 with five iterations, ours maintains competitive accuracy. The results suggest that our method, particularly with three iterations, offers superior enhancement compared to the other methods.

\begin{table}[h]
\caption{PHANTOM IMAGE RESULTS.}
\centering
\begin{tabular}{|c|c|c|c|c|c|}
  \hline
  Method&CNR 1&CNR 2&gCNR 1&gCNR 2 &SSIM\\
  \hline
  1 PW&3.6 db&1.4 db&0.9154&0.9352&0.72\\
  DANUIE\cite{Perdios}&5 db&4.4 db&0.9504&0.9358&0.77\\
  MSLAE\cite{Perdios2}&3.3 db&5 db&0.93&0.9525&o.76\\
  FCNN\cite{GOUDARZI}&6.1 db&3.3 db&0.9487&0.9496&0.76\\
  Ours (3 iters)&6.5 db&6.6 db&0.9305&0.9426&0.80\\
  Ours (5 iters)&6.5 db&8.5 db&0.9560&0.9629&0.81\\
  75 PW&8.5 db&8.1 db&0.9607&0.9429&1\\
  \hline
\end{tabular}
\label{TABLE3}
\end{table}

\begin{table}[h]
\caption{Results of NRMSE on the phantom data (RF).}
\centering
\begin{tabular}{|c|c|c|c|}
  \hline
  Method & INPUT& TARGET& NRMSE (output-target)\\
  \hline
   DANUIE \cite{Perdios} & 1 PW & 75 PW & 0.030 \\
  MSLAE \cite{Perdios2} & 1 PW & 75 PW & 0.027 \\
  FCNN \cite{GOUDARZI} & 1 PW & 75 PW & 0.030 \\
  Ours (3 iters) & 1 PW & 75 PW & 0.0196 \\
  Ours (5 iters) & 1 PW & 75 PW & 0.024 \\
  \hline
\end{tabular}
\label{TABLE4}
\end{table}

\subsubsection{Results on \textit{in vivo} data}
The efficacy of the proposed method is further assessed using carotid artery data to ascertain its ability to generalize the training performance to clinical applications. Visual comparisons with different methods are presented in Fig.~\ref{fig8}. Subfigures (e), (f), and (g) correspond to the outputs of the proposed method for 3, 5, and 7 iterations, respectively. All of those results eliminate clutter artifacts and enhance the visibility of the artery in the image. In contrast, when examining other methods, MSLAE\cite{Perdios2} achieves darkening in the vicinity of the artery, but some areas of the artery, as well as shallow portions of the resulting image, still exhibit artifacts (highlighted by red arrows) inherited from the input image (1 PW) depicted in Fig.~\ref{fig8} (a). DANUIE\cite{Perdios} method also has a lower level of enhancement than (b) but has the same issue with the artifacts. Finally, FCNN \cite{GOUDARZI} model (d) does not provide a  tangible enhancement in the image quality.

In terms of NRMSE, as documented in Table~\ref{TABLE5}, our method exhibits the closest output to the desired target compared to other methods. Notably, executing the reverse process by the proposed method for three iterations results in the closest approximation to the target. However, the similarity to the target diminishes with additional iterations. In contrast, for other methods, DANUIE \cite{Perdios} yields an NRMSE of 0.036. The MSLAE \cite{Perdios2} variant improves to an NRMSE of 0.034 under the same conditions. FCNN \cite{GOUDARZI}, akin to DANUIE \cite{Perdios}, achieves an NRMSE of 0.036. These results affirm that our method consistently outperforms other methods, even when subjected to seven iterations of the reverse process. Considering SSIM, as demonstrated in TABLE~\ref{TABLE5}, our proposed method achieves more index rather than the others with 0.81 and 0.80, respectively. In the speckle test, all of the methods have passed the test, meaning the speckle pattern has not been destroyed during the processing.

\begin{table}[h]
\caption{Results ON IN-VIVO DATA.}
\centering
\begin{tabular}{|c|c|c|c|c|c|}
  \hline
  Method & INPUT& TARGET& NRMSE&  SSIM &  KS test\\
  \hline
   DANUIE \cite{Perdios} & 1 PW & 75 PW & 0.030 & 0.74& Passed\\
  MSLAE \cite{Perdios2} & 1 PW & 75 PW & 0.027 & 0.75& Passed\\
  FCNN \cite{GOUDARZI} & 1 PW & 75 PW & 0.030 &0.75& Passed\\
  Ours (3 iters) & 1 PW & 75 PW & 0.0196 & 0.81& Passed \\
  Ours (5 iters) & 1 PW & 75 PW & 0.024& 0.80& Passed\\
  \hline
\end{tabular}
\label{TABLE5}
\end{table}

\section{Discussion}\label{Discussion}
Our proposed method demonstrates superior performance when compared to existing supervised deep learning-based models in the context of PW ultrasound denoising. A key aspect contributing to this efficacy lies in our approach of breaking down the noise into smaller, more manageable segments. This strategic division aims to facilitate a more targeted and efficient learning process for the model by simplifying the distribution learning task. By breaking down the noise, our method can effectively capture and address specific characteristics, leading to enhanced denoising outcomes.

We conducted our training using extensive RF images, adhering to the constraints of the Nyquist rate, thereby preventing any downsampling. While this approach ensures a high-fidelity representation of the ultrasound data, it introduces a computational burden due to the large size of the data. However, in future investigations, there is potential to explore the utilization of in-phase and quadrature (IQ) images, which offer more flexibility in terms of downsampling without violating the Nyquist rate constraints. This avenue could potentially streamline computations while maintaining the essential information required for effective PW ultrasound denoising.

The versatility of our formulation can be extended beyond its application on PW ultrasound denoising because we only trained the model using simulation data, and the method performed well in real phantom and in-vivo data. In other words, since generating simulated data is generally less expensive than performing real experiments, the method can be trained for other types of ultrasound imaging. This includes, but is not limited to, focused and synthetic aperture imaging, broadening the scope of its applicability across different ultrasound modalities. Moreover, our method is not confined to a specific type of ultrasound transducer; it can seamlessly be employed with diverse transducer configurations, such as convex and phased array transducers (granted that new simulated data is generated and the network is fine-tuned on the new data). This adaptability underscores the robustness and generalizability of our approach, making it a promising candidate for enhancing image quality and denoising capabilities across a spectrum of ultrasound imaging modalities and transducer types.

Deep learning models often show brittleness; although they provide good results for certain inputs, they can generate subpar results for others. Therefore, more experiments on simulated, real phantoms and in-vivo data are needed to ensure good performance in a large test database.

In future investigations, we plan to delve into the potential impact of the RF data denoised through our method on downstream tasks, with a specific focus on ultrasound elastography \cite{ashikuzzaman2024displacement,tehrani2022lateral}. This line of inquiry aims to explore whether the enhanced denoising capabilities of our method translate into improved performance and accuracy in subsequent higher-level applications, such as elastography.  In addition, reducing the processing time can be one of the prospective studies. The running time can be further improved
using knowledge distillation and pruning to achieve real-time performance.

While our method exhibits notable superiority over other approaches in the \textit{in vivo} test images, there remains a potential for further validation and improvement. Acknowledging this, our next research endeavor will focus on refining our network by incorporating additional \textit{in vivo} data. The intention is to fine-tune the model, leveraging an expanded dataset to enhance its adaptability and performance, specifically in real-world scenarios. This proactive approach to data collection and model refinement reflects our commitment to continuous improvement, ensuring that our method evolves to deliver even more impressive results in the dynamic and complex conditions encountered in \textit{in vivo} ultrasound imaging.

Similar to ~\cite{lee2023minimizing, song2020denoising} and~\cite{NEURIPS2023_80fe51a7}, the output of our model is not stochastic and is deterministic for a given input. In contrast to conventional diffusion models like ~\cite{10423849,chung2022come,li2023single}, which are trained  using only high-quality images, our method is trained on pairs of low/high-quality images to learn the process of transforming a low-quality image into a high-quality one. Therefore, if we start from a determined point (in our case, one PW) and pass through a deterministic process, knowing that the network does not have any randomness in the inference phase, the network returns the same output.

In this work, we did not use any stochasticity in the diffusion model and, therefore, did not take advantage of good mode convergence of these models. One idea for future work can be incorporating stochasticity with the proposed method model so that multiple plausible 75-PW compounded images can be reconstructed from a single sonification. This can ultimately help the clinicians using the technology by allowing inspection of different possible reconstructions. In other words, if different high-quality reconstructions are very different and can change clinical decision-making, the input of the diffusion-based reconstruction can be augmented with additional steering angles to narrow down the diverse set of modes, or alternatively, the full 75 steering angles can be collected to eliminate ambiguity.

\section{Conclusion}\label{Conclusion}
This paper proposes a denoising method in ultrasound PW imaging based on DDPMs. Our method was solely trained on a small dataset comprising of 400 simulated images, and performs well on simulation, real phantom as well as \textit{in vivo} data in the test phase. Diverging from traditional methodologies that simulate circular anatomies in data generation, our approach utilizes natural image segmentation masks as intensity maps for the simulated images. As a result, the proposed method showcases the ability to extend the denoising task to diverse anatomical shapes.

\begin{figure*}[t]
  \centering
  \includegraphics[width=1\textwidth]{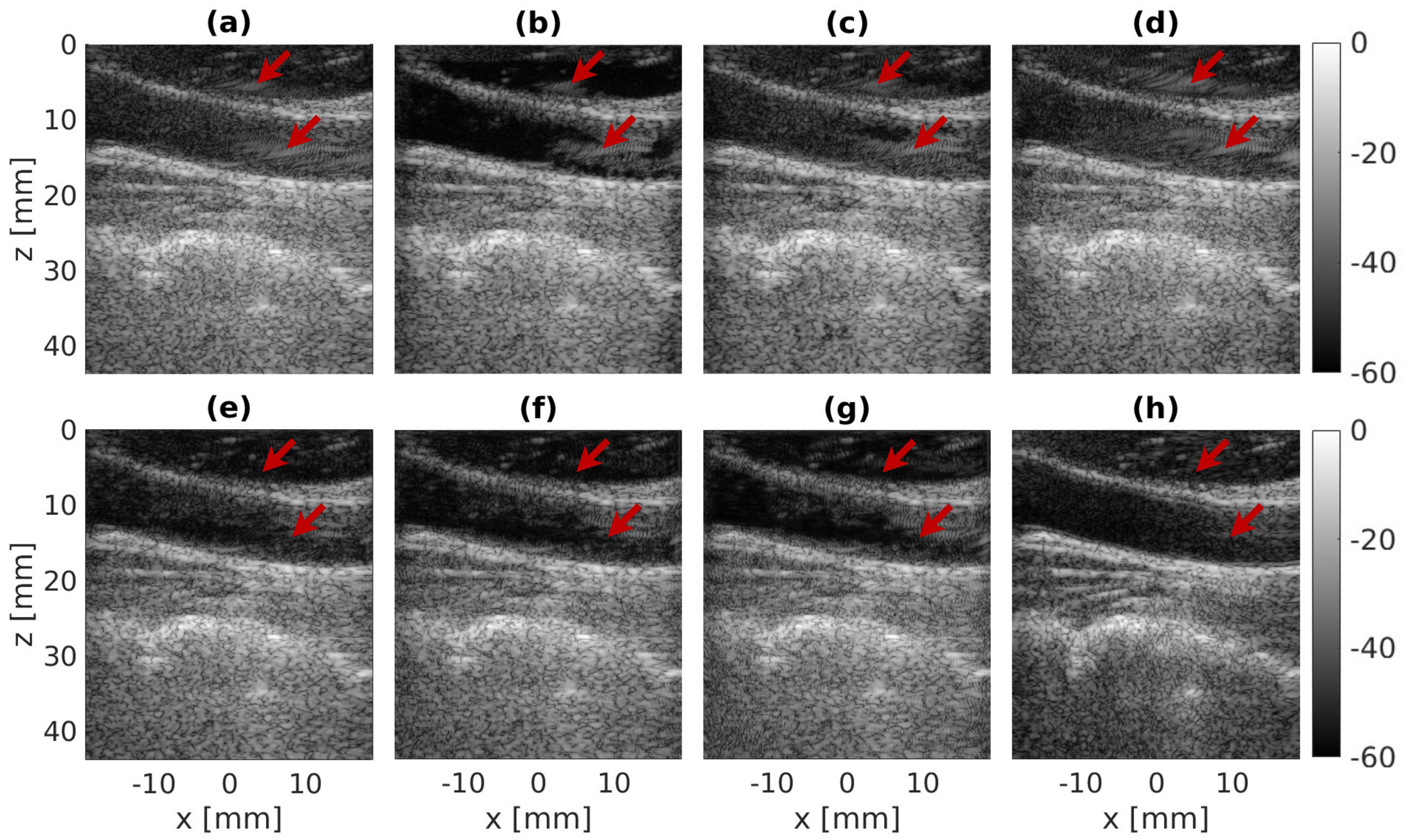}
  \caption{The \textit{in vivo} results for longitudinal view of the carotid artery. (a) Single PW image reconstructed using DAS. (b)  MSLAE\cite{Perdios2} result, (c) DANUIE \cite{Perdios} result, (d) FCNN \cite{GOUDARZI} result, (e) ours with 3 iterations, (f) ours with 5 iterations, (g) ours with 7 iterations (h) image reconstructed using 75 PWs.}
  \label{fig8}
\end{figure*}

\section*{Acknowledgement}
We would like to thank the anonymous reviewers for helping us improve the manuscript.



\bibliographystyle{ieeetr}

\end{document}